\documentstyle[preprint,aps]{revtex}

\begin{document}
\title{On the existence of a gyroscope in spaces with affine connections and metrics}
\author{S. Manoff\thanks{%
Permanent address: Bulgarian Academy of Sciences, Instutute for Nuclear
Research and Nuclear Energy, Department of Theoretical Physics, Blvd.
Tzarigradsko Chaussee 72, 1784 Sofia - Bulgaria}, B. Dimitrov}
\address{{\it Joint Institute for Nuclear Research,}\\
{\it Bogoliubov Laboratory of Theoretical Physics,}\\
{\it Dubna, Moscow Region}\\
{\it 141980 Russia}}
\date{e-mail addresses: {\it smanov@thsun1.jinr.ru, smanov@inrne.bas.bg,
bogdan@thsun1.jinr.ru}}
\maketitle

\begin{abstract}
Conditions for the existence of a gyroscope in spaces with affine
connections and metrics are found. They appear as special types of
Fermi-Walker transports for vector fields, lying in a subspace, orthogonal
to the velocity vector field (a non-null contravariant vector field) of an
observer.

\end{abstract}

\section{Introduction}

In the last years spaces with affine connections and metrics \cite{Norden} $%
\div $ \cite{Manoff-1} have deserved some interest related to the
possibility of using mathematical models of space-time different from
(pseudo) Riemannian spaces. The main reasons for new models of the
space-time are generated mostly by \cite{Hehl-1}:

(a) attempts to quantize gravity,

(b) attempts for description of ''hadron (or nuclear) matter in terms of
extended structure'' \cite{Hehl-1}, \cite{Hehl-2},

(c) attempts for modelling the early universe,

(d) models of continuous media with microstructure,

(e) models of classical and quantum field theories with more comprehensive
structure of the corresponding space-time.

The use of spaces with affine connections and metrics has been critically
evaluated from different points of view \cite{Hayashi}, \cite{Treder}. There
are at least four major objections against the applications of this types of
spaces in physics:

1. The violation of the equivalence principle related to the non-possibility
of bringing to zero the components of an affine connection on a curve in the
space-time,

2. The non-preservation (deformation) of a Lorentz basis along a geodesics
as a result of the non-compatibility of the affine connections (the parallel
transports) with the metrics (the measurement of lengths). This means that
in space with affine connections and metrics there are no transports along a
vector field preserving the lengths of vector fields and angles between
vector fields, transported along it,

3. The deformation of a light cone leading to the abuse of the law of
causality considered as a basic law in classical physics,

4. The independence of the affine connections and the metrics from each
other could lead to the determination of the affine connections in twofold
manner: on the one side, through the solution of some conditions for
compatibility [s. p. 2.] and, on the other side, through Lagrangian
formalism for both type of dynamic variables (the components of the affine
connections and the components of the metrics).

In the last few years the first three objections (1. $\div $ 3.) have been
removed by the investigations of different authors.

1a. It has been proved that in spaces with affine connections (whose
components differ only by sign or not only by sign) and metrics [i.e. in the
so called $(L_{n},g)$- and $(\overline{L}_{n},g)$-spaces] the principle of
equivalence holds \cite{Iliev-1} $\div $ \cite{Hartley}, \cite{Manoff-2}.

2a. In spaces with affine connections and metrics special types of
transports (called Fermi-Walker transports) \cite{Manoff-3} $\div $ \cite
{Manoff-5} exist which do not deform a Lorentz basis,

3c. There also exist other type of transports (called conformal transports) 
\cite{Manoff-6}, \cite{Manoff-6a} under which a light cone does not deform.

4a. The auto-parallel equation can play the same role in spaces with affine
connections and metrics for describing the motion of a free spinless test
particles as the geodesic equation does in Einstein's theory of gravitation 
\cite{Manoff-7}, \cite{Manoff-7a}.

\subsection{Problems and results}

The main purpose of the present paper is to show that the last objection 4.
could also be removed by means of the proof of the existence of a gyroscope
in spaces with affine connections and metrics. A gyroscope is characterized
by its three axes (in a $3$- or $4$-dimensional space-time) and the angles
between them. The length of the axes and the angles between them should not
change, when the gyroscope moves in the time or in the space-time. In this
sense a gyroscope represents a rigid body, determined by its axes \cite
{Synge}, \cite{Ehlers}. The existence of a gyroscope is related to the fact
that special types of Fermi-Walker transports could be found, which exist
for every preliminary given contravariant non-null vector field with its
corresponding projective metrics. For these types of transports the length
of the $n-1$ ($n=4$) gyroscope's axes lying in the $n-1$ dimensional
subspace could move in the time without changing their lengths and the
angles between them. The non-null contravariant vector field could be
interpreted as the velocity of an observer and the vectors, orthogonal to it
as the axes of a gyroscope. In this case, the independent to each other
affine connections and metrics would fulfill automatically compatibility
conditions for the special type of transports (related to the affine
connections) and the measurements of length (related to the metrics). The
affine connections and the metrics could be found uniquely by the use of a
Lagrangian formalism or by other methods common in physics.

Let us now consider the change of the length of a vector along a non-null
contravariant vector field in a space with affine connections and metrics.
The measuring of the length and its changes in space-time is very important
for theory and experiment in physics.

\section{Fermi-Walker transports (FWT) in subspaces with projective metrics}

Let a contravariant affine connection $\Gamma $ and a covariant affine
connection $P$ be given \cite{Manoff-1} with components in a co-ordinate
basis given respectively as $\Gamma _{jk}^{i}$ and $P_{jk}^{i}$ over a
differentiable manifold $M$ with $\dim M=n$. For $(L_{n},g)$-spaces $%
P_{jk}^{i}=-\Gamma _{jk}^{i}$. For $(\overline{L}_{n},g)$-spaces $%
P_{jk}^{i}+\Gamma _{jk}^{i}=g_{j;k}^{i}\neq 0$, where $g_{j}^{i}$ are the
components of the Kronecker tensor $Kr=g_{j}^{i}\cdot \partial _{i}\otimes
dx^{j}$. Let $M$ be provided with a covariant metric $g=g_{ij}\cdot
dx^{i}.dx^{j}$, $g_{ij}=g_{ji}$, $dx^{i}.dx^{j}=(1/2)\cdot (dx^{i}\otimes
dx^{j}+dx^{j}\otimes dx^{i})$ and its corresponding contravariant metric $%
\overline{g}=g^{ij}\cdot \partial _{i}.\partial _{j}$, $g^{ij}=g^{ji}$, $%
\partial _{i}.\partial _{j}=(1/2)\cdot (\partial _{i}\otimes \partial
_{j}+\partial _{j}\otimes \partial _{i})$. Let a non-null (non-isotropic)
vector field $u$ be given with $g(u,u)=e=\pm l_{u}^{2}\neq 0$, where $%
\overline{g}(g)(u)=\overline{g}[g(u)]=u$. The change of the length $l_{\xi
}=\mid g(\xi ,\xi )\mid ^{1/2}$ of another contravariant non-null vector
field $\xi $ along the vector field $u$ could be found in the form \cite
{Manoff-8} 
\begin{equation}
ul_{\xi }=\pm \frac{1}{2\cdot l_{\xi }}\cdot \lbrack (\nabla _{u}g)(\xi ,\xi
)+2\cdot g(\nabla _{u}\xi ,\xi )]\text{ , \ \ \ \ }l_{\xi }:\neq 0\text{ ,}
\label{1.1}
\end{equation}
where $ul_{\xi }=u^{i}\cdot \partial _{i}l_{\xi }=u^{i}\cdot (\partial
/\partial x^{i})l_{\xi }$, $\nabla _{u}\xi $ is the covariant derivative of $%
\xi $ along $u$, and $\nabla _{u}g$ is the covariant derivative of $g$ along 
$u$. Both covariant derivatives are with respect to the affine connections $%
\Gamma $ and $P$.

The change of the angle (the cosine of the angle respectively) between two
contravariant non-null vector fields $\xi $ and $\eta $ along the vector
field $u$ could be written in the form \cite{Manoff-8} 
\begin{eqnarray}
u[\cos (\xi ,\eta )] &=&\frac{1}{l_{\xi }\cdot l_{\eta }}\cdot \lbrack
(\nabla _{u}g)(\xi ,\eta )+g(\nabla _{u}\xi ,\eta )+g(\xi ,\nabla _{u}\eta
)]-  \nonumber \\
&&-[\frac{1}{l_{\xi }}\cdot (ul_{\xi })+\frac{1}{l_{\eta }}\cdot (ul_{\eta
})]\cdot \cos (\xi ,\eta )\text{ .}  \label{1.2}
\end{eqnarray}

The conditions for transports of the covariant metric $g$ and the conditions
for transports of the contravariant vector fields $\xi $ and $\eta $ as well
determine the change of the lengths of the two vector fields as well as the
angle between them.

To the vector field $u$ correspond its covariant projective metric $%
h_{u}=g-(1/e)\cdot g(u)\otimes g(u)$ and its contravariant projective metric 
$h^{u}=\overline{g}-(1/e)\cdot u\otimes u$.

If a vector field $\xi $ is orthogonal to the vector field $u$, i.e. if $%
g(u,\xi ):=0$, then $\xi $ could be written as $\xi _{\perp }=\overline{g}%
[h_{u}(\xi )]=g^{ij}\cdot h_{\overline{j}\overline{k}}\cdot \xi ^{k}\cdot
\partial _{i}$ in a $(\overline{L}_{n},g)$-space. The change $ul_{\xi
_{\perp }}$ of the length $l_{\xi _{\perp }}$ of $\xi _{\perp }$ could be
found as 
\begin{equation}
ul_{\xi _{\perp }}=\pm \frac{1}{2\cdot l_{\xi _{\perp }}}\cdot \lbrack
(\nabla _{u}h_{u})(\xi _{\perp },\xi _{\perp })+2\cdot h_{u}(\nabla _{u}\xi
_{\perp },\xi _{\perp })]\text{ ,}  \label{1.3}
\end{equation}
where 
\[
l_{\xi _{\perp }}^{2}=\pm h_{u}(\xi _{\perp },\xi _{\perp })\text{ , \ \ \ \
\ \ \ \ \ \ \ }l_{\xi _{\perp }}=\mid h_{u}(\xi _{\perp },\xi _{\perp })\mid
^{1/2}\text{ .} 
\]

The change of the angle between two orthogonal to $u$ vector fields $\xi
_{\perp }$ [with $g(u,\xi _{\perp })=0$] and $\eta _{\perp }$ [with $%
g(u,\eta _{\perp })=0$] could be computed and presented in the form 
\begin{eqnarray}
u[\cos (\xi _{\perp },\eta _{\perp })] &=&\frac{1}{l_{\xi _{\perp }}\cdot
l_{\eta _{\perp }}}\cdot \lbrack (\nabla _{u}h_{u})(\xi _{\perp },\eta
_{\perp })+h_{u}(\nabla _{u}\xi _{\perp },\eta _{\perp })+h_{u}(\xi _{\perp
},\nabla _{u}\eta _{\perp })]-  \nonumber \\
&&-[\frac{1}{l_{\xi _{\perp }}}\cdot (ul_{\xi _{\perp }})+\frac{1}{l_{\eta
_{\perp }}}\cdot (ul_{\eta _{\perp }})]\cdot \cos (\xi _{\perp },\eta
_{\perp })\text{ .}  \label{1.4}
\end{eqnarray}

The expressions for $ul_{\xi _{\perp }}$ and $u[\cos (\xi _{\perp },\eta
_{\perp })]$ contain only the covariant projective metric $h_{u}$, its
covariant derivative along $u$ and the \ corresponding vector fields $\xi
_{\perp }$ and $\eta _{\perp }$ as well as their covariant derivatives $%
\nabla _{u}\xi _{\perp }$, $\nabla _{u}\eta _{\perp }$, and the derivatives $%
ul_{\xi _{\perp }}$ and $ul_{\eta _{\perp }}$ along $u$.

The question arises under which conditions for $\nabla _{u}h_{u}$, $\nabla
_{u}\xi _{\perp }$, and $\nabla _{u}\eta _{\perp }$ the relations 
\begin{equation}
ul_{\xi _{\perp }}=0\text{, \ \ \ }ul_{\xi _{\perp }}=0\text{, \ \ }u[\cos
(\xi _{\perp },\eta _{\perp })]=0\text{ \ ,\ }  \label{1.5}
\end{equation}
are valid, i.e. under which conditions the lengths of the vector fields $\xi
_{\perp }$ and $\eta _{\perp }$, as well as the angle between them, do not
change under a transport along the vector field $u$. Transports which
preserve lengths and angles between vector fields are called Fermi-Walker
transports \cite{Manoff-3}, \cite{Manoff-6}. We can now apply the method,
developed for finding out Fermi-Walker transports in spaces with affine
connections and metrics with given metrics $g$ and $\overline{g}$, to the
same type of spaces with determined projective metrics $h_{u}$ and $h^{u}$.
This method [in details described in \cite{Manoff-5}] is related to the
introduction of an extended covariant differential operator $^{e}\nabla
_{u}=\nabla _{u}-\overline{A}_{u}$. The quantity $\overline{A}_{u}$ is a
tensor of the type 
\begin{equation}
\overline{A}_{u}=\overline{g}(C(u))=g^{ik}\cdot C_{\overline{k}j}(u)\cdot
\partial _{i}\otimes dx^{j}\text{ .}  \label{1.6}
\end{equation}

As a mixed tensor of second rank (depending on $u$) $\overline{A}_{u}$ has
to obey certain conditions under which \ if $^{e}\nabla _{u}\xi _{\perp
}=\nabla _{u}\xi _{\perp }-\overline{A}_{u}\xi _{\perp }=0$, i.e. if $\nabla
_{u}\xi _{\perp }=\overline{A}_{u}\xi _{\perp }=\overline{g}(C(u))(\xi
_{\perp })$, then $ul_{\xi _{\perp }}=0$, \ \ \ $ul_{\xi _{\perp }}=0$, and
\ \ $u[\cos (\xi _{\perp },\eta _{\perp })]=0$.

Using the relation $\overline{g}=h^{u}+(1/e)\cdot u\otimes u$, we can
represent $\nabla _{u}\xi _{\perp }=\overline{A}_{u}\xi _{\perp }$ in the
form 
\begin{equation}
\nabla _{u}\xi _{\perp }=h^{u}(C(u))(\xi _{\perp })+\frac{1}{e}\cdot \lbrack
(u)(C(u))](\xi _{\perp })\cdot u\text{.}  \label{1.7}
\end{equation}

After introducing the last expression in the relation for $ul_{\xi _{\perp
}} $ and after some calculations we can find the relations 
\begin{equation}
ul_{\xi _{\perp }}=\pm \frac{1}{2\cdot l_{\xi _{\perp }}}\cdot \{(\nabla
_{u}h_{u})(\xi _{\perp },\xi _{\perp })+2\cdot h_{u}[h^{u}(C(u))(\xi _{\perp
}),\xi _{\perp }]\}\text{ \ \ ,}  \label{1.8}
\end{equation}
\begin{equation}
\lbrack h_{u}(h^{u})(C(u))]_{s}=-\frac{1}{2}\cdot \nabla _{u}h_{u}\text{ \ ,}
\label{1.9}
\end{equation}
where 
\begin{eqnarray}
h_{u}(h^{u})(C(u)) &=&h_{ik}\cdot h^{\overline{k}\overline{l}}\cdot
C_{kj}(u)\cdot dx^{i}\otimes dx^{j}=  \nonumber \\
&=&h^{\overline{k}\overline{l}}\cdot h_{ik}\cdot C_{kj}(u)\cdot
dx^{i}\otimes dx^{j}=h^{u}[h_{u}\otimes C(u)]\text{ \ ,}  \nonumber \\
h_{u}[h^{u}(C(u))(\xi _{\perp }),\xi _{\perp }] &=&\{h^{u}[h_{u}\otimes
C(u)]\}(\xi _{\perp },\xi _{\perp })\text{ ,}  \label{1.10} \\
\lbrack h_{u}(h^{u})(C(u))]_{s} &=&\frac{1}{2}\cdot \{h^{u}[h_{u}\otimes
C(u)]+h^{u}[C(u)\otimes h_{u})]\}\text{ ,}  \nonumber \\
\lbrack h_{u}(h^{u})(C(u))]_{a} &=&\frac{1}{2}\cdot \{h^{u}[h_{u}\otimes
C(u)]-h^{u}[C(u)\otimes h_{u})]\}\text{ .}  \nonumber
\end{eqnarray}

If (\ref{1.9}) is fulfilled, then $ul_{\xi _{\perp }}=0$. Therefore, to
every Fermi-Walker transport with a given tensor $C(u)$ there exists a
corresponding Fermi-Walker transport for the orthogonal to the vector field $%
u$ contravariant vector fields $\{\xi _{\perp }\in T^{\perp u}(M)\}$. On the
other side, to every given tensor $C(u)$, i.e. to every given extended
covariant differential operator $^{e}\nabla _{u}=\nabla _{u}-\overline{g}%
(C(u))$, there exists a Fermi-Walker transport for the orthogonal to $u$
contravariant vector fields $\{\xi _{\perp }\in T^{\perp u}(M)\}$. A
Fermi-Walker transport of this type is described by the condition (\ref{1.9}%
) or in a co-ordinate basis by the condition 
\begin{equation}
h_{ik}\cdot h^{\overline{k}\overline{l}}\cdot C_{lj}(u)+h_{jk}\cdot h^{%
\overline{k}\overline{l}}.C_{li}(u)=-h_{ij;k}\cdot u^{k}\text{ \ .}
\label{1.11}
\end{equation}

Since \cite{Manoff-5} $C_{lj}(u)=(A_{lj\overline{k}}+B_{lj;k})\cdot u^{k}$,
we have for every arbitrary given non-null vector field $u$ the conditions
for $h_{ij}$%
\begin{equation}
h_{ij;k}=-[h_{ik}\cdot h^{\overline{k}\overline{l}}\cdot (A_{lj\overline{k}%
}+B_{lj;k})+h_{jk}\cdot h^{\overline{k}\overline{l}}\cdot (A_{li\overline{k}%
}+B_{li;k})]\text{ ,}  \label{1.11a}
\end{equation}
where $A_{lj\overline{k}}$ and $B_{lj;k}$ are the components of tensor
fields determining the different types of Fermi-Walker transports in spaces
with affine connections and metrics for given metrics $g$ and $\overline{g}$.

Taking into account the preservation of the lengths $l_{\xi _{\perp }}$ and $%
l_{\eta _{\perp }}$ under the above conditions for a FWT we can find the
remaining conditions for the preservation of the angle between $l_{\xi
_{\perp }}$ and $l_{\eta _{\perp }}$ under their transport along the vector
field $u$. For $u[\cos (\xi _{\perp },\eta _{\perp })]$ we obtain [if $%
ul_{\xi _{\perp }}=0$, $ul_{\xi _{\perp }}=0$] 
\begin{eqnarray}
u[\cos (\xi _{\perp },\eta _{\perp })] &=&\frac{1}{l_{\xi _{\perp }}\cdot
l_{\eta _{\perp }}}\cdot \{(\nabla _{u}h_{u})(\xi _{\perp },\eta _{\perp
})+h_{u}[h^{u}(C(u))(\xi _{\perp }),\eta _{\perp }]+  \nonumber \\
&&+h_{u}[h^{u}(C(u))(\eta _{\perp }),\xi _{\perp }]\}\text{ \ .}
\label{1.12}
\end{eqnarray}

From the relations 
\begin{eqnarray}
(\nabla _{u}h_{u})(\xi _{\perp },\eta _{\perp }) &=&-2\cdot \lbrack
h_{u}(h^{u})(C(u))]_{s}(\xi _{\perp },\eta _{\perp })\text{ ,}  \nonumber \\
h_{u}[h^{u}(C(u))(\xi _{\perp }),\eta _{\perp }]
&=&[h_{u}(h^{u})(C(u))](\eta _{\perp },\xi _{\perp })\text{ ,}  \nonumber \\
h_{u}[h^{u}(C(u))(\eta _{\perp }),\xi _{\perp }] &=&[h_{u}(h^{u})(C(u))](\xi
_{\perp },\eta _{\perp })\text{ \ ,}  \label{1.13}
\end{eqnarray}
after representing the last two expression in their symmetric and
antisymmetric parts, it follows that $u[\cos (\xi _{\perp },\eta _{\perp
})]=0$ is automatically fulfilled.

Therefore, in spaces with affine connections and metrics a gyroscope could
exist if its axes $\xi _{\perp }^{(b)}$ [$b=1$, $...$ , $n-1$; $n=3,4,...$]
are transported under a Fermi-Walker transport along a worldline with a
tangent vector $u$. The FWT for the axes of the gyroscope is determined by
the condition 
\begin{eqnarray}
h_{u}(\nabla _{u}\xi _{\perp }^{(b)}) &=&-\frac{1}{2}\cdot (\nabla
_{u}h_{u})(\xi _{\perp }^{(b)})+[h_{u}(h^{u})(C(u))]_{a}(\xi _{\perp
}^{(b)})=  \nonumber \\
&=&-\frac{1}{2}\cdot (\nabla _{u}h_{u})(\xi _{\perp
}^{(b)})+\{h^{u}[h_{u}\otimes C(u)]\}_{a}(\xi _{\perp }^{(b)})\text{ .}
\label{1.14}
\end{eqnarray}

If we chose $\xi _{\perp }^{(b)}$ in a way that $\pounds _{u}\xi _{\perp
}^{(b)}=0$ (i.e. if $u$ and $\xi _{\perp }^{(b)}$ are tangent vectors to the
co-ordinates in the space-time $M$), then \cite{Manoff-9}, \cite{Manoff-10} 
\begin{eqnarray}
h_{u}(\nabla _{u}\xi _{\perp }^{(b)}) &=&(\sigma +\frac{1}{n-1}\cdot \theta
\cdot h_{u})(\xi _{\perp }^{(b)})+\omega (\xi _{\perp }^{(b)})=  \nonumber \\
&=&[h_{u}(h^{u})(C(u))]_{s}(\xi _{\perp
}^{(b)})+[h_{u}(h^{u})(C(u))]_{a}(\xi _{\perp }^{(b)})\text{ ,}  \label{1.15}
\end{eqnarray}
where \cite{Manoff-9} the tensor $\sigma $ is the shear velocity tensor
(shear), the invariant $\theta $ is the expansion velocity invariant
(expansion), and $\omega $ is the rotation velocity tensor (rotation). From
the last two expressions (for every arbitrary given vector $\xi _{\perp }$
on which the covariant differential operator $h_{u}\circ \nabla _{u}$ acts)
we obtain the relations 
\begin{eqnarray}
\lbrack h_{u}(h^{u})(C(u))]_{s} &=&\sigma +\frac{1}{n-1}\cdot \theta \cdot
h_{u}\text{ ,}  \nonumber \\
\lbrack h_{u}(h^{u})(C(u))]_{a} &=&\omega \text{ .}  \label{1.16}
\end{eqnarray}

For a Fermi-Walker transport for $\xi _{\perp }$ along $u$, we have the
condition in the form 
\begin{equation}
\nabla _{u}h_{u}=-2\cdot (\sigma +\frac{1}{n-1}\cdot \theta \cdot h_{u})%
\text{ .}  \label{1.17}
\end{equation}

\subsection{Fermi-Walker transports along a shear-free and / or
expansion-free vector field $u$}

For a shear-free vector field $u$ ($\sigma =0$) the condition (\ref{1.17})
for a FWT degenerates in the recurrent relation for $h_{u}$%
\begin{equation}
\nabla _{u}h_{u}=-\frac{2}{n-1}\cdot \theta \cdot h_{u}\text{ , \ \ \ \ \ \
\ \ \ }\pounds _{u}\xi _{\perp }=0\text{ .}  \label{1.18}
\end{equation}

For a shear-free and expansion-free vector field $u$ ($\sigma =0$, $\theta
=0 $) \cite{Manoff-7a} the condition (\ref{1.17}) for a FWT degenerates in
the condition for a parallel transport of $h_{u}$ along $u$%
\begin{equation}
\nabla _{u}h_{u}=0\text{ .}  \label{1.19}
\end{equation}

In this type of spaces with affine connections and metrics the parallel
transport of two contravariant non-null vector fields $\xi _{\perp }$ and $%
\eta _{\perp }$ along the vector field $u$ assure the preservation of its
lengths and angles between them.

\subsection{Some remarks}

1. If a (pseudo) Riemannian space (as a special case of a space with affine
connections and metrics) admits a Killing vector field $u$, i.e. if $\pounds
_{u}g=0$ , leading to $\sigma =\theta =0$, then a FWT is determined by the
parallel transport of $h_{u}$.

2. The conditions for the existing of a gyroscope in spaces with affine
connections and metrics do not determine the affine connections and the
metrics. For every space with given affine connections and metrics there
exist Fermi-Walker transports preserving the axes of \ a gyroscope and the
angles between them. The conditions for a special FWT determine only the
transport of the vector fields $\xi _{\perp }$, and $\eta _{\perp }$ along a
given vector field $u$ in a given space-time. The compatibility between
metrics (measurements of lengths) and affine connections (transports which
are not geodesic) is automatically fulfilled by the choice of the
corresponding transport preserving lengths and angles. Thus, the last
objections for using spaces with affine connections and metrics as models of
space-time is removed.

\section{Conclusions}

In the present paper special types of Fermi-Walker transports are consider
under which a gyroscope can exist in spaces with affine connections and
metrics. These types of transports are in general different from the
geodesic transports but they play the same role as the geodesic transports
in (pseudo) Riemannian spaces and, at the same time, do not put any
conditions on the affine connections and metrics. The last two geometric
objects could be determined by other methods uniquely since the
compatibility conditions between them are fulfilled automatically under the
special types of Fermi-Walker transports. Physical theories (including
theories of gravitation) could be constructed in the above mentioned types
of spaces not only for microphysics (quantum physics) but also for
macrophysics (classical physics).

\end{document}